%% file: arx_met.tex
\include{smaria_def}

\documentclass[prl,twocolumn,groupedaddress]{revtex4}
\usepackage[letterpaper]{geometry}
\usepackage{amsfonts}
\usepackage{graphicx}
\usepackage{psfrag}
\begin{document}
\begin{flushleft}
\vspace*{-1.0in}
\hspace*{1.25in} 
\mbox EFI-01-22,FNAL-PUB-01/084-E \\ %
\end{flushleft}
\bibliographystyle{apsrev}
\title[lala] {Search for Gluinos and Scalar Quarks in $p\bar{p}$ Collisions
at $\sqrt{s}=1.8$ TeV using the Missing Energy plus
Multijets Signature}
\date{\today}
\begin{abstract}
We have performed a search for gluinos ($\gls$) and squarks ($\sq$) in
a data sample of \mbox{84 $\rm{pb}^{-1}~$} of \ppb~ collisions at
$\sqrt{\mbox{s}}$ = 1.8 TeV, recorded by the Collider Detector at
Fermilab, by investigating the final state of large missing transverse
energy and 3 or more jets, a characteristic signature in
$R$-parity-conserving supersymmetric models. The analysis has been
performed `blind', in that the inspection of the signal region is made
only after the predictions from Standard Model backgrounds have been
calculated.  Comparing the data with predictions of constrained
supersymmetric models, we exclude gluino masses below \mbox{195 \gev}
(95\% C.L.), independent of the squark mass.  For the case $\msq
\approx \mgls$, gluino masses below \mbox{300 \gev} are excluded.\\

PACS number(s):14.80.Ly, 12.60.-i, 12.60.Jv, 13.85.Rm, 05.45.Jn
\end{abstract}
\maketitle
\font\eightit=cmti8
\def\r#1{\ignorespaces $^{#1}$}
\hfilneg
\begin{sloppypar}
\noindent
T.~Affolder,\r {23} H.~Akimoto,\r {45}
A.~Akopian,\r {37} M.~G.~Albrow,\r {11} P.~Amaral,\r 8  
D.~Amidei,\r {25} K.~Anikeev,\r {24} J.~Antos,\r 1 
G.~Apollinari,\r {11} T.~Arisawa,\r {45}
% ...
\end{sloppypar}
\vskip .026in
\begin{center}
(CDF Collaboration)
\end{center}

\vskip .026in
\begin{center}
\r 1  {\eightit Institute of Physics, Academia Sinica, Taipei, Taiwan 11529, 
Republic of China} \\
\r 2  {\eightit Argonne National Laboratory, Argonne, Illinois 60439} \\
\r 3  {\eightit Istituto Nazionale di Fisica Nucleare, University of Bologna,
I-40127 Bologna, Italy} \\
\r 4  {\eightit Brandeis University, Waltham, Massachusetts 02254} \\
\r 5  {\eightit University of California at Davis, Davis, California  95616} \\
\r 6  {\eightit University of California at Los Angeles, Los 
Angeles, California  90024} \\  
\r 7  {\eightit Instituto de Fisica de Cantabria, CSIC-University of Cantabria, 
39005 Santander, Spain} \\
\r 8  {\eightit Enrico Fermi Institute, University of Chicago, Chicago, 
Illinois 60637} \\
\r 9  {\eightit Joint Institute for Nuclear Research, RU-141980 Dubna, Russia}
\\
\r {10} {\eightit Duke University, Durham, North Carolina  27708} \\
\r {11} {\eightit Fermi National Accelerator Laboratory, Batavia, Illinois 
60510} \\
\r {12} {\eightit University of Florida, Gainesville, Florida  32611} \\
\r {13} {\eightit Laboratori Nazionali di Frascati, Istituto Nazionale di Fisica
               Nucleare, I-00044 Frascati, Italy} \\
\r {14} {\eightit University of Geneva, CH-1211 Geneva 4, Switzerland} \\
\r {15} {\eightit Glasgow University, Glasgow G12 8QQ, United Kingdom}\\
\r {16} {\eightit Harvard University, Cambridge, Massachusetts 02138} \\
\r {17} {\eightit Hiroshima University, Higashi-Hiroshima 724, Japan} \\
\r {18} {\eightit University of Illinois, Urbana, Illinois 61801} \\
\r {19} {\eightit The Johns Hopkins University, Baltimore, Maryland 21218} \\
\r {20} {\eightit Institut f\"{u}r Experimentelle Kernphysik, 
Universit\"{a}t Karlsruhe, 76128 Karlsruhe, Germany} \\
\r {21} {\eightit Center for High Energy Physics: Kyungpook National
University, Taegu 702-701; Seoul National University, Seoul 151-742; and
SungKyunKwan University, Suwon 440-746; Korea} \\
\r {22} {\eightit High Energy Accelerator Research Organization (KEK), Tsukuba, 
Ibaraki 305, Japan} \\
\r {23} {\eightit Ernest Orlando Lawrence Berkeley National Laboratory, 
Berkeley, California 94720} \\
\r {24} {\eightit Massachusetts Institute of Technology, Cambridge,
Massachusetts  02139} \\   
\r {25} {\eightit University of Michigan, Ann Arbor, Michigan 48109} \\
\r {26} {\eightit Michigan State University, East Lansing, Michigan  48824} \\
\r {27} {\eightit University of New Mexico, Albuquerque, New Mexico 87131} \\
\r {28} {\eightit The Ohio State University, Columbus, Ohio  43210} \\
\r {29} {\eightit Osaka City University, Osaka 588, Japan} \\
\r {30} {\eightit University of Oxford, Oxford OX1 3RH, United Kingdom} \\
\r {31} {\eightit Universita di Padova, Istituto Nazionale di Fisica 
          Nucleare, Sezione di Padova, I-35131 Padova, Italy} \\
\r {32} {\eightit University of Pennsylvania, Philadelphia, 
        Pennsylvania 19104} \\   
\r {33} {\eightit Istituto Nazionale di Fisica Nucleare, University and Scuola
               Normale Superiore of Pisa, I-56100 Pisa, Italy} \\
\r {34} {\eightit University of Pittsburgh, Pittsburgh, Pennsylvania 15260} \\
\r {35} {\eightit Purdue University, West Lafayette, Indiana 47907} \\
\r {36} {\eightit University of Rochester, Rochester, New York 14627} \\
\r {37} {\eightit Rockefeller University, New York, New York 10021} \\
\r {38} {\eightit Rutgers University, Piscataway, New Jersey 08855} \\
\r {39} {\eightit Texas A\&M University, College Station, Texas 77843} \\
\r {40} {\eightit Texas Tech University, Lubbock, Texas 79409} \\
\r {41} {\eightit Institute of Particle Physics, University of Toronto, Toronto
M5S 1A7, Canada} \\
\r {42} {\eightit Istituto Nazionale di Fisica Nucleare, University of Trieste/
Udine, Italy} \\
\r {43} {\eightit University of Tsukuba, Tsukuba, Ibaraki 305, Japan} \\
\r {44} {\eightit Tufts University, Medford, Massachusetts 02155} \\
\r {45} {\eightit Waseda University, Tokyo 169, Japan} \\
\r {46} {\eightit University of Wisconsin, Madison, Wisconsin 53706} \\
\r {47} {\eightit Yale University, New Haven, Connecticut 06520} \\
\r {(\ast)} {\eightit Now at Carnegie Mellon University, Pittsburgh,
Pennsylvania  15213} \\
\r {(\ast\ast)} {\eightit Now at Northwestern University, Evanston, Illinois 
60208}
\end{center}
\collaboration{CDF}
\affiliation{\r 1  {\eightit Institute of Physics, Academia Sinica, Taipei, Taiwan 11529, 
Republic of China} \\
\r 2  {\eightit Argonne National Laboratory, Argonne, Illinois 60439} \\
\r 3  {\eightit Istituto Nazionale di Fisica Nucleare, University of Bologna,
I-40127 Bologna, Italy} \\
\r 4  {\eightit Brandeis University, Waltham, Massachusetts 02254} \\
\r 5  {\eightit University of California at Los Angeles, Los 
Angeles, California  90024} \\  
\r 6  {\eightit Instituto de Fisica de Cantabria, University of Cantabria, 
39005 Santander, Spain} \\
\r 7  {\eightit Enrico Fermi Institute, University of Chicago, Chicago, 
Illinois 60637} \\
\r 8  {\eightit Joint Institute for Nuclear Research, RU-141980 Dubna, Russia}
\\
\r 9  {\eightit Duke University, Durham, North Carolina  27708} \\
\r {10}  {\eightit Fermi National Accelerator Laboratory, Batavia, Illinois 
60510} \\
\r {11} {\eightit University of Florida, Gainesville, Florida  32611} \\
\r {12} {\eightit Laboratori Nazionali di Frascati, Istituto Nazionale di Fisica
               Nucleare, I-00044 Frascati, Italy} \\
\r {13} {\eightit University of Geneva, CH-1211 Geneva 4, Switzerland} \\
\r {14} {\eightit Harvard University, Cambridge, Massachusetts 02138} \\
\r {15} {\eightit Hiroshima University, Higashi-Hiroshima 724, Japan} \\
\r {16} {\eightit University of Illinois, Urbana, Illinois 61801} \\
\r {17} {\eightit The Johns Hopkins University, Baltimore, Maryland 21218} \\
\r {18} {\eightit Institut f\"{u}r Experimentelle Kernphysik, 
Universit\"{a}t Karlsruhe, 76128 Karlsruhe, Germany} \\
\r {19} {\eightit Korean Hadron Collider Laboratory: Kyungpook National
University, Taegu 702-701; Seoul National University, Seoul 151-742; and
SungKyunKwan University, Suwon 440-746; Korea} \\
\r {20} {\eightit High Energy Accelerator Research Organization (KEK), Tsukuba, 
Ibaraki 305, Japan} \\
\r {21} {\eightit Ernest Orlando Lawrence Berkeley National Laboratory, 
Berkeley, California 94720} \\
\r {22} {\eightit Massachusetts Institute of Technology, Cambridge,
Massachusetts  02139} \\   
\r {23} {\eightit Institute of Particle Physics: McGill University, Montreal 
H3A 2T8; and University of Toronto, Toronto M5S 1A7; Canada} \\
\r {24} {\eightit University of Michigan, Ann Arbor, Michigan 48109} \\
\r {25} {\eightit Michigan State University, East Lansing, Michigan  48824} \\
\r {26} {\eightit University of New Mexico, Albuquerque, New Mexico 87131} \\
\r {27} {\eightit The Ohio State University, Columbus, Ohio  43210} \\
\r {28} {\eightit Osaka City University, Osaka 588, Japan} \\
\r {29} {\eightit University of Oxford, Oxford OX1 3RH, United Kingdom} \\
\r {30} {\eightit Universita di Padova, Istituto Nazionale di Fisica 
          Nucleare, Sezione di Padova, I-35131 Padova, Italy} \\
\r {31} {\eightit University of Pennsylvania, Philadelphia, 
        Pennsylvania 19104} \\   
\r {32} {\eightit Istituto Nazionale di Fisica Nucleare, University and Scuola
               Normale Superiore of Pisa, I-56100 Pisa, Italy} \\
\r {33} {\eightit University of Pittsburgh, Pittsburgh, Pennsylvania 15260} \\
\r {34} {\eightit Purdue University, West Lafayette, Indiana 47907} \\
\r {35} {\eightit University of Rochester, Rochester, New York 14627} \\
\r {36} {\eightit Rockefeller University, New York, New York 10021} \\
\r {37} {\eightit Rutgers University, Piscataway, New Jersey 08855} \\
\r {38} {\eightit Texas A\&M University, College Station, Texas 77843} \\
\r {39} {\eightit Texas Tech University, Lubbock, Texas 79409} \\
\r {40} {\eightit Istituto Nazionale di Fisica Nucleare, University of Trieste/
Udine, Italy} \\
\r {41} {\eightit University of Tsukuba, Tsukuba, Ibaraki 305, Japan} \\
\r {42} {\eightit Tufts University, Medford, Massachusetts 02155} \\
\r {43} {\eightit Waseda University, Tokyo 169, Japan} \\
\r {44} {\eightit University of Wisconsin, Madison, Wisconsin 53706} \\
\r {45} {\eightit Yale University, New Haven, Connecticut 06520} \\}

The ~\sm (SM) ~\cite{GSW} 
accurately describes physical phenomena down to
scales of $\sim 10^{-16}~\rm cm$.  There are many extensions of the
\sm~ to smaller length scales, including extra gauge interactions, new
matter, new levels of compositeness, and supersymmetry (SUSY).  
Of these, supersymmetry~\cite{coleman} treats the bosonic and
fermionic degrees of freedom equally and provides a robust 
extension to the \sm~. 
For simplicity the {\it minimal} 
construction (MSSM) is often used to link SUSY with the \sm~\cite{MSSM}.  
The most general MSSM would induce proton decay with a
weak-interaction lifetime; to avoid this, baryon and lepton
conservation are enforced in the MSSM by postulating a new conserved
quantity, $R$-parity, $R=(-1)^{3(B-L) + 2 s}$, where for each particle
$s$ is the
spin, and $B$ and $L$ are the respective baryon and lepton
assignments. $R$-parity conservation leads to
characteristic SUSY signatures with missing transverse energy in the
final state due to the stable lightest supersymmetric particle (LSP).
We assume in the search described below for the bosonic partners of
quarks (squarks) and the fermionic partners of gluons (gluinos) that
the LSP is weakly interacting, as is the case for most of the MSSM
parameter space.

We consider gluino and squark production within the minimal
supergravity model (mSUGRA)\cite{MSSM}.  In this model 
the entire SUSY mass
spectrum is essentially determined by only five unknown parameters:
the common scalar mass at the GUT scale, $M_0$; the common gaugino
mass at the GUT scale, $M_{1/2}$; the common trilinear coupling at the
GUT scale, $A_{0}$; the sign of the Higgsino mixing parameter, 
$sign(\mu)$; 
and the ratio of the Higgs vacuum expectation values, $\tan\beta$.  
Minimal SUGRA does not make predictions for the part of
the $\msq$-$\mgls$ mass parameter space where squarks of the first two
families are lighter than about 0.8 times the mass of the gluino.  
Hence for $\msq < \mgls$ we use the constrained MSSM~\cite{MSSM} 
with the set of input parameters being  the
mass of the gluino, $\mgls$;
the $CP$-odd neutral scalar Higgs mass, $m_{A}$; the squark masses,
$\msq_{i}$; the slepton masses, $m_{\stilde{\ell_{i}}}$; the
squark and slepton mixing parameters, $A_{t(b)(\tau)}$ ; and
$\mu$ and $\tan\beta$.   

We investigate whether the production and decay of gluinos and 
scalar quarks  is observable in the rate of ~$\ge$3-jet events
with large missing transverse energy  at the Collider
Detector at Fermilab (CDF).   
The large missing energy would originate from the two LSPs 
in the final states of the squark and gluino 
decays. The three or more  hadronic jets  would result 
from the hadronic decays of the $\sq$ and/or $\gls$.
We use the \isaj\ Monte Carlo (MC) 
program ~\cite{isajet} with $\tan\beta=3$ to generate datasets of
squark and gluino events, and the {\sc prospino} program ~\cite{prospino} to
calculate the production cross sections. To be conservative, only the
first two generations of squarks ($\tilde{u},\tilde{d},\tilde{c},\tilde{s})$
are assumed to be produced~\cite{stop_mixing} in the general MSSM
framework; we additionally consider production of the bottom squark
($\tilde{b}$) in the mSUGRA case. 
The search is based on \mbox{$84 \pm 4$ pb$^{-1}$} of
integrated luminosity recorded with the CDF detector
during the 1994-95 Tevatron run.

The CDF detector is described in detail elsewhere~\cite{cdf_d}.
The momenta of charged particles
are measured in the central tracking chamber (CTC), which 
is positioned inside a 1.4 T superconducting solenoidal magnet. 
Outside the magnet, electromagnetic and
hadronic calorimeters arranged in a projective tower geometry cover the
pseudorapidity region \mbox{$|\eta| < 4.2$~\cite{CDFcoo}} and are used
to identify jets. Jets are defined as localized energy depositions 
in the calorimeters and are reconstructed 
using an iterative clustering algorithm with a fixed cone of radius
\mbox{$\rm{\it{\Delta R}} \equiv \sqrt{\rm{\it{\Delta}}\eta^2 + \rm{\it{\Delta}}\phi^2} = 0.7$} in 
$\eta - \phi$ space ~\cite{jets}. 
Jets are ordered in transverse
energy, $E_T=E\sin\theta$, where $E$ is the scalar sum of energy 
deposited in the calorimeter towers within the cone, and $\theta$ is
the angle formed by the beam-line, the event vertex~\cite{multiple_vertices}, 
and the cone center. 

The missing transverse energy is defined as the negative vector sum of the
transverse energy in the electromagnetic and hadronic calorimeters,
\mbox{$\vecmet = -\sum_{i} (E_{i}\sin\theta_{i})\hat{n}_{i}$}, where $E_{i}$
is the energy of the {\it i}-th tower, $\hat{n}_{i}$ is a transverse
unit vector pointing to the center of each tower, and $\theta_{i}$ is
the polar angle of the tower; the sum extends to $|\eta|<3.6$. 
The data sample was selected with an on-line trigger which 
requires $\met \equiv |\vecmet| > 30$ GeV.

We use a two-stage preselection to reject accelerator- and
detector-related backgrounds, beam halo, and cosmic ray events.  The
first stage is based on timing and energy information in the
calorimeter towers to reject events out-of-time with a $\ppbar$
collision. The second stage uses the event electromagnetic
fraction (${F}_{em}$) and event charged fraction (${F}_{ch}$) to
distinguish between real and fake jet events~\cite{eemf}.  The
preselection requirements and the corresponding missing transverse
energy spectra are presented in Figure 1.
\begin{figure}
\label{fig1}
\includegraphics[width=3.2in]{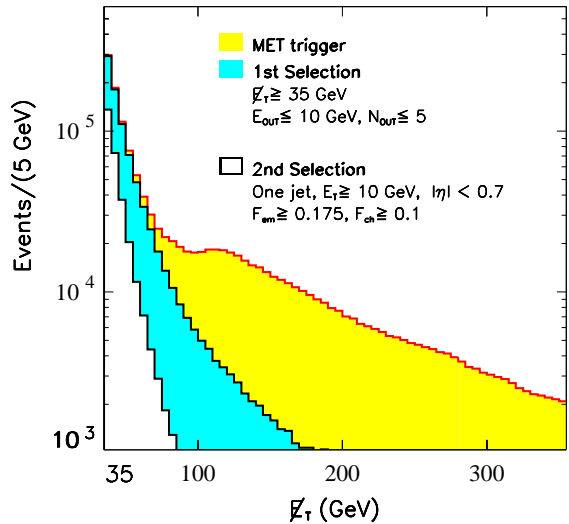}
\caption{ The \met~spectrum after the online trigger ~\protect{\cite{beletini}}
and the two stages of the data preselection. The numbers 
of events surviving the
first and second selections are 892,395 and 286,728, respectively. The
variables $E_{OUT},~N_{OUT}$ are energy and number of towers 
out of time~\protect{\cite{smaria_thesis}}. }
\end{figure}
At least three jets with $E_T\ge 15$ GeV, at least one of them within
$|\eta|< 1.1$, are then required in events that pass the
preselection.  A total of 107,509 events, predominantly from QCD
multijet production, survive the three-jet requirement.

The observed missing energy in QCD jet production is largely
a result of jet mismeasurements and
detector resolution. 
In a QCD multijet event with large missing energy, 
the highest \et\ jet is typically the most 
accurately measured.  When the second 
or third jet is mismeasured because it lands 
partially in  an uninstrumented region (a `gap'),
the \met\ is pulled close in $\phi$ to the mismeasured 
jet.
A jet is considered non-fiducial if it is within 0.5 rad in
$\phi$ of the \met\ direction and also points in $\eta$ to a detector
gap.  The second and third highest \et\ jets in an event are
required to be fiducial.  We eliminate the residual QCD component by
using the correlation in the $\delta\phi_{1}=|\phi_{{\rm leading
~jet}}-\phi_{\met}|$ versus $\delta\phi_{2}= |\phi_{{\rm second~
jet}}-\phi_{\met}|$ plane.  We accept events with $R_{1} > 0.75$ rad
and $R_{2} > 0.5$ rad, where
$R_{1}=\sqrt{\delta\phi_{2}^{2}+(\pi-\delta\phi_{1})^{2}}
{\rm and}~ R_{2}=\sqrt{\delta\phi_{1}^{2}+(\pi-\delta\phi_{2})^{2}}$.

To avoid potential {\it a posteriori} biases when searching for new
physics in the tails of the missing transverse energy distribution, once we
define the signal candidate data sample we make it inaccessible. This
analysis approach is often referred to as a `blind analysis' and the
signal candidate data sample as a `blind box'. The `blind box' data
are inspected only after the entire search path has been defined by
estimating the total \sm\ backgrounds and
optimizing the sensitivity to the supersymmetric
signal. We use three variables to define the signal candidate region :
\met, $\Ht \equiv E_{T(2)}+E_{T(3)}+\met$, and isolated track
multiplicity, \niso ~\cite{niso}.  The `blind box' contains events
with $\met\ge 70$ GeV, \mbox{$H_T\ge 150$ GeV}, and \niso=0.  The large
missing transverse energy requirement for 
the definition of the box is motivated
by the requirement that the trigger be fully efficient
\cite{smaria_thesis}. The \Ht\ requirement provides good
discrimination between signal and
background~\cite{smaria_thesis}.
The \niso\ requirement increases the
sensitivity of the search for all-hadronic final states by
significantly reducing the backgrounds from $W/Z$+jets and top-antitop
(\ttb) events while retaining the signal cascade decays in which a
lepton is produced close to a jet (non-isolated lepton).  The analysis
path is shown in Table \ref{tab2}.
\begin{table}
\caption{The data selection path for the \met~+$\ge$ 3 jets
search. After the fourth step, all events that could fall 
in the `blind box' are removed from the accounting. 
The events tabulated in the following steps are only in the control bins.}  	
\label{tab2}
\begin{ruledtabular}
\begin{tabular}{ll} 
{Requirement}&{Events}\\ \hline
{Preselection}&{286,728 }\\
{$N_{jet}\ge$3 ($\rm{\it{\Delta R}} =0.7$, \et$\ge$ 15 GeV) }&{107,509}\\ \hline
{Fiducial 2nd, 3rd jet}& {57,011}\\ \hline
{$R_{1}>$0.75 rad,$R_{2}>$0.5 rad}&{23,381} \\ \hline\hline 
{\met$\ge$70 GeV, \Ht $\ge$150 GeV,}&{}\\
{\niso=0}& {`blind box'}\\ \hline\hline
{$E_{T(1)} \ge$70GeV}&{ } \\
{$E_{T(2)} \ge$30GeV}&{ }\\
{$|\eta|(1~or~2~or~3)<1.1$} & {6435}\\ \hline 
{$f_{em(1)},~f_{em(2)}\le$ 0.9}&{6013} \\ \hline
{L2 trigger requirement}&{4679}\\ \hline
{$\delta\phi_{min}(\met-jet)\ge 0.3$ rad}&{2737}\\ 
\end{tabular}
\end{ruledtabular}
\end{table}
We reduce the background contribution from $W(\rightarrow e\nu)+$jets
and \ttb\ production by requiring the two highest energy jets not be
purely electromagnetic (jet electromagnetic fraction $f_{em}<$0.9).
We further reduce the contribution from QCD backgrounds (mismeasured
jets) by requiring the \met\ vector not be closer than 0.3 rad in
$\phi$ to any jet in the event.

We estimate the $W$ and $Z$ boson backgrounds by using a leading order
perturbative QCD calculation for $W(Z)+$ jets as implemented in the
\vecb\ Monte Carlo \cite{vecbos}, enhanced with a coherent parton
shower evolution of both initial- and final-state partons,
hadronization, and a soft underlying event model
(\vecb+\herw~\cite{herwig}). 
Events with large missing transverse energy and
$\ge$3  jets in the final state are expected primarily from
$Z(\rightarrow \nu\bar{\nu})+\ge$3 jets and $W(\rightarrow
\tau\nu)+\ge$2 jets (the third jet originating from the hadronic
$\tau$ decay) processes.  The MC predictions for events with $\ge3$ jets
are normalized to the observed $Z (\to ee)+$ jets data sample
via the measured $N_{jet}\over{N_{jet}+1}$ ratio, where $N_{jet}$ is
the number of jets.  The ratio 
$\rho\equiv\frac{\sigma(\ppb \rightarrow W(\rightarrow e\nu)+jets)}{
\sigma(\ppb\rightarrow Z(\rightarrow e^{+}e^{-})+jets)}$ is used
to normalize the $W$ MC predictions.  Assuming lepton universality,
the predictions for the number of events with $\ge 2$-- and $\ge
3$--jets from $W$ and $Z$ production and decay to all flavors are
normalized to the data for $Z(\rightarrow e^{+}e^{-})+\ge 2$ jets.  By
normalizing the MC predictions to data we avoid large systematic
effects due to the renormalization scale, the choice of parton density
functions, initial- and final-state radiation, and the jet energy
scale.  The total uncertainty ($\sim$10\%) is then dominated by the
uncertainty on the luminosity measurement, the uncertainty on the
measured ratio $N_{jet}\over{N_{jet}+1}$, and the
uncertainty on the predicted ratio $\rho$ as a function of $N_{jet}$.

We estimate the backgrounds from single top, \ttb, and diboson events
with Monte Carlo predictions normalized  using
the respective theoretical cross section calculations for these processes.
We generate \ttb\ events with the \pyt\ MC
program \cite{pythia}, normalizing to  the fully resummed theoretical cross
section $\sigma_{\ttb}=5.06^{+0.13}_{-0.36}$ pb for $m_{top}=175 {\rm
~GeV}/c^2$ \cite{tdib}. We assign a total uncertainty of $\pm 18\%$ on the
cross-section to take into account the uncertainty  
on the top quark mass.
The top quark can also be produced singly 
via $W$-gluon fusion and \qqb\
annihilation with cross sections of \mbox{$\sigma_{Wg}=1.7$ pb ($\pm$
17\%)}, and
\mbox{$\sigma_{W^*\rightarrow t\bar{b}}=0.73$ pb($\pm$ 9\%)} 
~\cite{tdib}.  We use the \herw\ \cite{herwig}\ ($W$-gluon fusion) 
and \pyt\ (\qqb\ annihilation) programs to generate the 
single top production processes.  
We generate boson pair production with the \pyt\ MC and use the
calculated cross sections
\mbox{$\sigma_{WW}=9.5\pm 0.7$ pb}, \mbox{$\sigma_{WZ}=2.6\pm 0.3$ pb} and 
\mbox{$\sigma_{ZZ}=1.0 \pm 0.2$ pb} \cite{tdib}. 

The data samples we use to study and normalize the QCD Monte Carlo
predictions consist of events collected by on-line identification of
at least one jet with transverse energy above trigger thresholds of 20
and 50 GeV, and with integrated luminosity of
\mbox{0.094 pb$^{-1}$} and 2.35 pb$^{-1}$, respectively. The corresponding 
QCD MC samples are generated using the \herw~ program and 
a CDF detector simulation. The shapes of the \met\ and jet multiplicity
distributions are in good agreement with the data, as are the
jet kinematic distributions.  The QCD predictions are absolutely normalized
to the data for $N_{jet}\ge 3$. The total uncertainty on the QCD
background estimate is $\sim$15\%, dominated by a 12\%  uncertainty due to
the detector resolution.

There are seven bins around the `blind box' formed by inverting the
requirements which define it ({\it i.e.} by changing the direction of
the inequalities shown in the bin definitions of Table
~\ref{tab3}).  
We compare the \sm\ background predictions in the bins
around the `blind box' with the data. The results are shown in Table
~\ref{tab3}. 
\begin{table}
\caption{Comparison of the ~\sm~prediction and the data in the
bins neighboring bin 8, the `blind box'. After the contents of cons 
were compared in detail to standard model predictions, 
we `opened the box'. We find 74 events in bin 8.}  	
\label{tab3}
\begin{ruledtabular}
\begin{tabular}{c|c|c|c|c}
Bin Definition&EWK&QCD&All&Data\\ \hline 
{\met~$\ge$70,\Ht$\ge$150,\niso$>0$}&{14}&{6.3}&{20$\pm$5}&{10}\\ \hline
{\met~$\ge$70,\Ht$<$150,\niso$=0$}&{2.3}&{6.3}&{8.6$\pm$4.5}&{12}\\ \hline
{35$<$\met~$<$70,\Ht$>$150,\niso$=0$}&{1.95}&{135}&{137$\pm$28}&{134}\\ \hline
{\met~$>$70,\Ht$<$150,\niso$>0$}&{1.73}&{$<$0.1}&{1.73$\pm$0.3}&{2}\\ \hline
{35$<$\met~$<$70,\Ht$>$150,\niso$>0$}&{14}&{9.4}&{23.4$\pm$6}&{24}\\ \hline
{35$<$\met~$<$70,\Ht$<$150,\niso$=0$}&{5}&{413}&{418$\pm$69}&{410}\\ \hline
{35$<$\met~$<$70,\Ht$<$150,\niso$>0$}&{3.3}&{28}&{31$\pm$10}&{35}\\ \hline
{\met~$\ge$70,\Ht$\ge$150,\niso$=0$}&{35}&{41}&{76$\pm$13}&{$\sqcap$}\\ 
\end{tabular}
\end{ruledtabular}
\end{table}
Of the 35 events from electroweak processes 
predicted in the `blind box', $\sim$37\% are expected from
$Z\rightarrow \nu\bar{\nu}+\ge 3$ jets, $\sim$20\% from
$W\rightarrow \tau\nu+\ge 2$ jets, $\sim$20\% from the combined
$W\rightarrow e(\mu)\nu_{e}(\nu_{\mu})+\ge 3$ jets, and $\sim$20\%
from \ttb\ production and decays. We also compare the kinematic
properties between \sm\ predictions and the data around the box and 
find them to be in agreement ~\cite{smaria_thesis}.

To probe
the SUSY parameter space in a simple and comprehensive way
we divide the $\msq-\mgls$ plane into four general regions : (A)
$\msq>\mgls$ (mSUGRA, five $\sq$); (B) $\msq\sim\mgls$ (mSUGRA,
five $\sq$); (C) $\msq<\mgls$ (MSSM, four $\sq$); (D)
$\msq\ll\mgls$ (MSSM, four $\sq$).  We analyze representative
points of each region and optimize the \met\ and \Ht\ requirements for
increased sensitivity to the signal using MC data. The ratio
$N_{SUSY}\over\sqrt{N_{SM}}$ is maximized 
in region A for \met~$\ge 90$ GeV and \Ht$\ge$ 160 GeV; in region B for
\met~$\ge 110$ and \Ht$\ge$ 230 GeV; in C for \met~$\ge 110$
and \Ht$\ge$ 170 GeV; and in D for \met~$\ge 90$ and \Ht$\ge$ 160 GeV,
where $N_{SUSY}$ is the number of signal events and $N_{SM}$ is the
number of \sm\ background events. The signal efficiency ranges between
1\% and 14\% for the different points in the parameter space, and its
total relative systematic uncertainty (mostly due to parton density
functions, gluon radiation, renormalization scale and jet energy
scale) ranges between 10\% and 15\%.

In the `blind box', where we expect 76$\pm$13 \sm\ events, we observe
74 events.  In Figure 2 the predicted \sm\ kinematic
distributions are compared with the distributions we observe in the
data.  For the A/D, B and C region requirements, we observe 31, 5 and
14 events where we expect 33 $\pm~7$, 3.7 $\pm ~0.5$ and 10.6 $\pm$ ~0.9
events respectively.  
\begin{figure}
\label{fig2}
\includegraphics[width=3.2in]{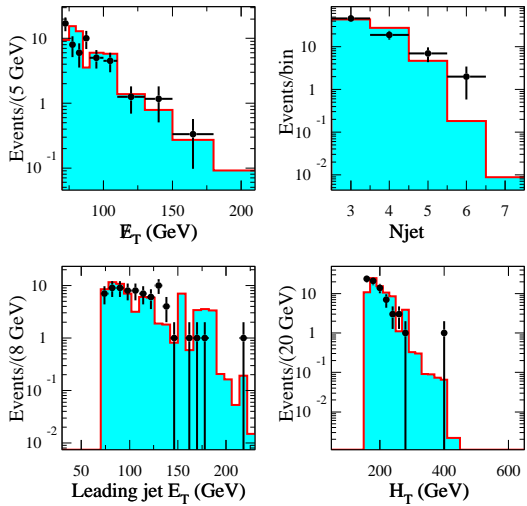}
\caption{ Comparison in the `blind box' 
between data (points) and \sm\ predictions (histogram) of ~\met,
$N_{jet}$, leading jet \et\ and \Ht\ distributions. There are 74
events in each of these plots, to be compared with 76$\pm$13 SM
predicted events. Note that the \met\ distribution is plotted with
a variable bin size; the bin contents are normalized as labelled.}
\end{figure}
Based on the observations, the \sm\ estimates and their uncertainties,
and the relative total systematic uncertainty on the signal
efficiency, we derive the 95\% C.L. ~\cite{lim} upper limit on the
number of signal events. The bound is shown on the $\msq-\mgls$ plane in
Figure 3.  For the signal points generated with mSUGRA the limit is
also interpreted in the $M_{0}-M_{1/2}$ plane ~\cite{smaria_thesis}.
Studies of the dependence on the value of $\tan\beta$ can be found in
\cite{msug_wrkshp,tripler}.   
\begin{figure}
\label{fig3}
\includegraphics[width=3.2in]{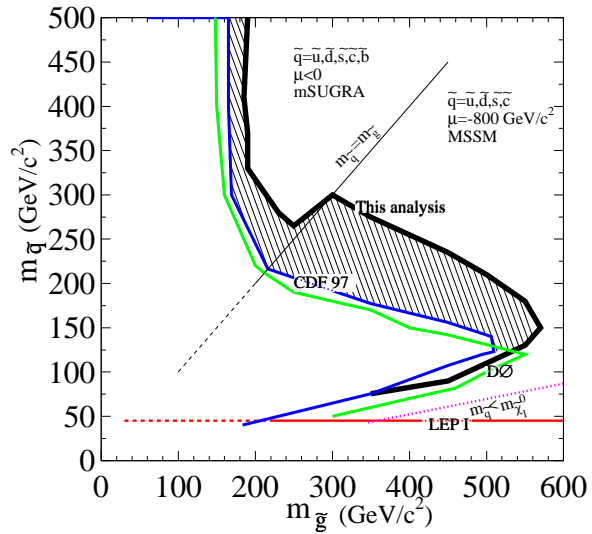}
\caption{The 95\% C.L. limit curve in the $\msq~-~\mgls$ plane 
for $\tan\beta=3$; the hatched area is newly excluded by this
analysis. Results from some previous searches are also
shown(CDF~\protect\cite{Abe:1997}, D\O ~\protect\cite{Abachi:1995ng},
LEP I~\protect\cite{Abreu:1990he}; the area at lower masses in the
plane has been previously excluded by the UA1 and UA2
experiments~\protect\cite{Albajar:1987yu, Alitti:1990ux}.
The region labelled as $\msq < m_{\stilde{\chi}_{1}^{0}}$ is
theoretically forbidden as the squarks are predicted to be lighter
than the LSP.}
\end{figure}

In conclusion, a search for gluinos and squarks in events with large missing
energy plus multijets excludes at 95\% C.L.
gluino masses below 300 \gev~ for the case $\msq \approx \mgls$, and
below 195 \gev, independent of the squark mass, in constrained
supersymmetric models. This is a significant
extension of previous bounds. 

We thank the Fermilab staff and the technical staffs of the
participating institutions for their vital contributions.  This work was
supported by the U.S. Department of Energy and National Science Foundation;
the Italian Istituto Nazionale di Fisica Nucleare; the Ministry of Education,
Science, Sports and Culture of Japan; the Natural Sciences and Engineering 
Research Council of Canada; the National Science Council of the Republic of 
China; the Swiss National Science Foundation; the A. P. Sloan Foundation; the
Bundesministerium fuer Bildung und Forschung, Germany; the Korea Science 
and Engineering Foundation (KoSEF), the Korea Research Foundation, and the 
Comision Interministerial de Ciencia y Tecnologia, Spain.

\par

\end{document}

%% file: smaria_def.tex
\def\ap#1,#2,#3#4{           {\it Ann. Phys. (NY)\/ }{\bf #1} #2 (19#3#4)}
\def\apj#1,#2,#3#4{          {\it Astrophys. J.\/ }{\bf #1} #2 (19#3#4)}
\def\apjl#1,#2,#3#4{         {\it Astrophys. J. Lett.\/ }{\bf #1} #2 (19#3#4)}
\def\app#1,#2,#3#4{          {\it Acta Phys. Polon.\/ }{\bf #1} #2 (19#3#4)}
\def\com#1,#2,#3#4{          {\it Comm. Math. Phys.\/ }{\bf #1} #2 (19#3#4)}
\def\ib#1,#2,#3#4{           {\it ibid.\/ }{\bf #1} #2 (19#3#4)}
\def\nat#1,#2,#3#4{          {\it Nature (London)\/ }{\bf #1} #2 (19#3#4)}
\def\np#1,#2,#3#4{           {\it Nucl. Phys.\/ }{\bf B#1} #2 (19#3#4)}
\def\npps#1,#2,#3#4{         {\it Nucl. Phys. B (Proc. Suppl.)\/ }{\bf B#1}
                             #2 (19#3#4)}
\def\plb#1,#2,#3#4{          {\it Phys. Lett.\/ }{\bf B#1} #2 (19#3#4)}
\def\pla#1,#2,#3#4{          {\it Phys. Lett.\/ }{\bf A#1} #2 (19#3#4)}
\def\prd#1,#2,#3#4{          {\it Phys. Rev.\/ }{\bf D#1} #2 (19#3#4)}
\def\prep#1,#2,#3#4{         {\it Phys. Rep.\/ }{\bf #1} #2 (19#3#4)}
\def\prl#1,#2,#3#4{          {\it Phys. Rev. Lett.\/ }{\bf #1} #2 (19#3#4)}
\def\pro#1,#2,#3#4{          {\it Prog. Theor. Phys.\/ }{\bf #1} #2 (19#3#4)}
\def\rmp#1,#2,#3#4{          {\it Rev. Mod. Phys.\/ }{\bf #1} #2 (19#3#4)}
\def\sp#1,#2,#3#4{           {\it Sov. Phys.-Usp.\/ }{\bf #1} #2 (19#3#4)}

\def\zp#1,#2,#3#4{           {\it Zeit. fur Physik\/ }{\bf C#1} #2 (19#3#4)}
\def\vecmet{\mbox{$\vec{\not\!\!{E}}_T$}}

\def\vecb{{\sc vecbos}}

\def\isaj{{\sc  isajet}}
\def\herw{{\sc  herwig}}

\def\pyt{{\sc  pythia}}

\def\ttb{\mbox{$t\bar{t}$}}

\def\ppb{\mbox{$p\bar{p}$}}
\def\qqb{\mbox{$q\bar{q}$}}

\def\vereq#1#2{\lower3pt\vbox{\baselineskip1.5pt \lineskip1.5pt}}
\def\lesssim{\mathrel{\mathpalette\vereq<}}
\def\gtrsim{\mathrel{\mathpalette\vereq>}}
\def\beq{\begin{equation}}
\newcommand{\Ht}{\mbox{$H_{T}$}}
\newcommand{\et}{\mbox{$E_{T}$}}
\newcommand{\pt}{\mbox{$P_{T}$}}

\newcommand{\bspace}{\!\!\!\!}
\newcommand{\met}{\mbox{$E{\bspace}/_{T}$}}
\newcommand{\niso}{\mbox{$N_{trk}^{iso}$}}

\newcommand{\ppbar}{\mbox{$p\overline{p}$}}

\newcommand{\gev}{\mbox{GeV/$c^2$}}

\def\stilde{\widetilde}
\newcommand{\sq}{\stilde{q}}

\newcommand{\gls}{\stilde{g}}
\newcommand{\msq}{\mbox {$m_{\stilde{q}}$} }
\newcommand{\mgls}{\mbox {$m_{\stilde{g}}$} }

\def\eeq{\end{equation}}
\def\bea{\begin{eqnarray}}
\def\beaa{\begin{eqnarray*}}
\def\eea{\end{eqnarray}}
\def\eeaa{\end{eqnarray*}}
\def\bq{\begin{quote}}
\def\eq{\end{quote}}

\def\gappeq{\mathrel{\rlap {\raise.5ex\hbox{$>$}}
{\lower.5ex\hbox{$\sim$}}}}
\def\lappeq{\mathrel{\rlap{\raise.5ex\hbox{$<$}}
{\lower.5ex\hbox{$\sim$}}}}
\def\sm{Standard Model}

%% file: arx_met.bbl
\begin{references}
\bibitem{multiple_vertices} If there are multiple vertices in the
event we use the vertex with the largest $\sum p_{T}$ of associated tracks.
\bibitem{eemf}
${F}_{em}= \frac{\sum_{j=1}^{N_{jet}}\et_{j}\times 
f_{em(j)}}{\sum_{j=1}^{N_{jet}}\et_{j}}$, where $N_{jet}$ is the number of
jets of cone $0.7$ with \et$>10$ GeV and  $f_{em(j)}$ is the Electromagnetic
fraction of the  $j$-th jet. ${F}_{ch} =
\left<\frac{(\sum_{i}^{tracks}\pt_{i})_{j}}{\et_{j}}\right>$  where $(\sum_{i}^{tracks}\pt_{i})_{j}$ is the sum of the \pt~ of all the tracks $i$ matched with a central jet $j$.
\bibitem{beletini} The shoulder in the online \met\ distribution is
due to the contribution of a trigger that requires a jet above 100 GeV.
\end{references}
